\documentclass{article}
\usepackage[T1]{fontenc} 
\usepackage[utf8]{inputenc} 
\usepackage{ismir,amsmath,amsfonts,cite,url}
\usepackage{graphicx}
\usepackage{color}
\usepackage{booktabs}

\title{Jazz Contrafact Detection}

\twoauthors
  {Carey Bunks} {City, University of London}
  {Tillman Weyde} {City, University of London}

\def\authorname{C. Bunks and T. Weyde}

\usepackage[bookmarks=false,pdfauthor={\authorname},pdfsubject={\papersubject},hidelinks]{hyperref}

\begin{document}

\maketitle
\begin{abstract}
In jazz, a contrafact is a new melody composed over an existing, but often reharmonized chord progression. Because reharmonization can introduce a wide range of variations, detecting contrafacts is a challenging task. This paper develops a novel vector-space model to represent chord progressions, and uses it for contrafact detection. The process applies principles from music theory to reduce the dimensionality of chord space, determine a common key signature representation, and compute a chordal co-occurrence matrix.  The rows of the matrix form a basis for the vector space in which chord progressions are represented as piecewise linear functions, and harmonic similarity is evaluated by computing the \emph{membrane area}, a novel distance metric.  To illustrate our method's effectiveness, we apply it to the Impro-Visor corpus of 2,612 chord progressions, and present examples demonstrating its ability to account for reharmonizations and find contrafacts.
\end{abstract}

\section{Introduction}
\label{sec:introduction}

In jazz, a contrafact is a song whose harmony is similar to that of another song, but has a different melody \cite{kernfeld2002new}. The tune \emph{Rhythm Changes}, by George Gershwin (1930),
is a well-known source of many contrafacts \cite{wiki2022rhythmchanges}, and there are numerous examples of other jazz standards that also have them \cite{wiki2018list}. 

In addition to the difference in melody, contrafact chord progressions often have reharmonizations, a common practice in jazz that makes chord substitutions in a song while maintaining its harmonic identity \cite{berkman2013jazz}. Reharmonization is a core characteristic of jazz -- so much so that it is not unusual for different published versions of the same song to exhibit significant variations. Furthermore, jazz musicians regularly employ reharmonization as an improvisational technique in both live and studio-recorded performances.  

These observations suggest that the harmonic similarity challenge encountered in contrafact detection will also play an important role in music information retrieval tasks such as cover song identification \cite{schedl2014music}.  The analysis of harmonic similarity has been studied by others, for example, using parse trees and hierarchical models \cite{de2012music, rohrmeier2020syntax}.  The approach taken in this paper, however, is a novel method that accounts for reharmonizations using distributional semantics \cite{boleda2020distributional}. The implementation generates a vector space embedding for chords based on a co-occurrence matrix \cite{lund1996producing}, computed from a corpus of symbolic jazz chord progressions.

As many chords in our corpus occur only rarely, the dimensionality of chord space must be reduced \cite{bruni2014multimodal}. Typically this would be accomplished after vectorization, using truncated SVD or principal component analysis \cite{sorzano2014survey}. 
In this paper, however, we propose a new approach: we use principles of music theory to reduce the dimensionality of chord space before vector embedding. 

The ensuing sections provide a comprehensive review of the source data, followed by an in-depth description of how we apply music theory to achieve a reduction in the dimensionality of chord space.  The characteristics of the resulting co-occurrence matrix are discussed, and a novel distance metric, the \emph{membrane area}, is used for evaluating harmonic similarity and detecting contrafacts.

\section{The Data}
The data used in this paper is a corpus of symbolic chord progressions similar to those found in jazz fake books, such as the Real Book \cite{leonard2016real}.  
The progressions are mainly jazz standards, but also include some blues, jazz-blues, modal jazz, and pop tunes. 
The corpus is derived from a collection distributed with \emph{Impro-Visor}, an open-source music notation program intended to help musicians learn improvisation \cite{keller2008Impro}.  Our modifications remove control information used by the Impro-Visor application, retaining the musical content.  

The Impro-Visor corpus provides chord progressions for 2,612 songs, and is the largest digital collection of its type that we know of.  The applications iRealPro\footnote{\url{https://www.irealb.com/forums/showthread.php?12753-Jazz-1350-Standards}} and Band-in-a-Box\footnote{\url{https://members.learnjazzstandards.com/sp/biab-jazzstandards/}} contain chord progressions for roughly 1400 and 226 jazz standards, respectively.  The Weimer Jazz Database contains chords for 456 jazz songs (along with transcribed solos and extensive annotations).\footnote{\url{https://jazzomat.hfm-weimar.de/dbformat/dbcontent.html}}  

The Impro-Visor corpus contains 134,182 chord symbols, and it is a rich collection, consisting of 1,542 unique types. If the frequency of chord symbols were uniform, there would be 87 instances of each one in the corpus. In fact, 20\% of the chord symbols occur just a single time, and 50\% fewer than six times.  

As the corpus consists mainly of jazz standards, it contains a preponderance of $7^{th}$ chords, comprising of at least a root, $3^{rd}$, $5^{th}$, and $7^{th}$ notes. 
These types of chords often have additional extensions ($9^{th}$, $11^{th}$, $13^{th}$) and chromatic alterations ($\flat$9, $\sharp$9, $\flat$5, $\sharp$5). 
A common variation of jazz chords replaces the $7^{th}$ with a $6^{th}$. 
As $7^{th}$ chords are the basic harmonic unit in jazz \cite{rawlins2005jazzology}, 
and make up 77\% of our corpus, they are the focus of our approach to dimensionality reduction described in the next section.

Of the remaining chords, 16\%  (21,860) are three-note chords (triads), and  7\% are drawn from a variety of special types.  
These latter include \emph{slash} chords, \emph{sus} chords, \emph{power} chords, \emph{polychords}, major triads with an added 9$^{th}$, and the \emph{no-chord} symbol. 
Every chord in our corpus belongs to one of these categories, as listed in Table~\ref{tab:chordclass}, along with their frequencies. 

\begin{table}
\centering
\begin{tabular}{l|r|r}
\textbf{Type}   &
\textbf{Number} & \textbf{Percentage} \\
\hline \hline
major7           & 19448           & 14.494\%            \\
dominant7        & 46944           & 34.985\%            \\
minor7           & 30471           & 22.709\%            \\
minor7$\flat$5         & 3568            & 2.659\%             \\
diminished7      & 2807            & 2.092\%             \\
major triad      & 15410           & 11.484\%            \\
major triad add9 & 170             & 0.127\%             \\
minor triad      & 5796            & 4.320\%             \\
diminished triad & 128             & 0.095\%             \\
augmented triad  & 526             & 0.392\%             \\
slash chord      & 6415            & 4.781\%             \\
sus chord        & 1830            & 1.364\%             \\
no chord         & 615             & 0.458\%             \\
power chord      & 42              & 0.031\%             \\
polychord        & 12              & 0.009\%             \\
\hline
\textbf{Totals}  & \textbf{134182} & \textbf{100\%}     
\end{tabular}
\caption{Corpus chord types and their frequencies}
\label{tab:chordclass}
\end{table}

\section{Dimensionality Reduction}
\label{sec:modelorder}

As discussed in the previous section, 20\% of the chord symbols in the corpus occur only a single time and 50\% fewer than six times.  A co-occurrence matrix based on one-hot chord vectors of dimension 1,542 would be sparse and ill-conditioned, suggesting the need for dimensionality reduction. The issue of infrequent symbol types is well known in natural language processing (NLP).  NLP handles dimensionality reduction with techniques such as principal component analysis, truncated singular value decomposition, and gradient descent, applied after having mapped words to vector embeddings \cite{sorzano2014survey}. 

This paper takes a different approach.   
We apply principles of music theory to reduce the dimensionality of chord space prior to vector embedding. Specifically, we draw on chord-scale theory as described in \cite{mulholland2013berklee} to select a set of 60 chord classes, 
based on 5 chord types and 12 scale positions 
to represent the core characteristics of jazz 
harmony.
In the following sections, we specify a method for mapping each chord in the corpus to one of these classes, accounting for all 1,542 types. 
There are many ways that music theory can be used to describe chord relationships. 
In the following, we describe the details and rationale of our specific approach.

\subsection{$7^{th}$ Chord Types}\label{subsec:diatonic}

\begin{figure}
 \centerline{\includegraphics[width=0.975\columnwidth]
 {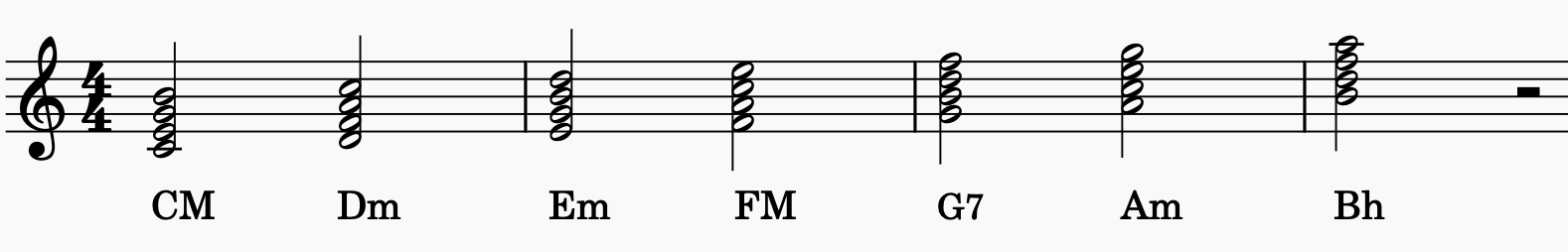}
}
 \caption{Diatonic $7^{th}$ chords of the C major scale}
 \label{fig:diatonic}
\end{figure}

Our choice of chord  types focuses on the four-note $7^{th}$ chords diatonically generated from the 12 major scales.  Figure~\ref{fig:diatonic} shows the chords obtained from the C major scale, 
illustrating that
there are four chord types: 
major7 (M), minor7 (m), dominant7 (7), and minor7$\flat$5 (h), where the symbols shown in parentheses are abbreviations we use in this paper.

Each chord type can be represented  
by its defining chord, as well as versions of this chord with any extensions or alterations.  
For example, we map chords such as Cm9 and Cm11 to the Cm7 class; C7$\flat$9, C7$\sharp$5, and C13 map to the C7 class; and CM7$\sharp$11 maps to the CM7 class.  
In addition, in accordance with reharmonization practices, we assign chords such as CmM7 to the Cm7 class and C6 to the CM7 class.

In addition to chords diatonic to major scales, we add one additional type: the diminished7 (o) chord. Although not generated by major scales, the diminished7 plays an important and unique role in jazz \cite{rawlins2005jazzology, levine2011jazz}. 

Finally, we include an additional class to account for the \emph{no-chord} symbol.  
This symbol is used in the corpus to designate a momentary absence of harmony, and is important to include as a metric spacer when comparing the harmonic similarity of two songs. 

\subsection{Chord Type Mapping}
The first five rows of Table~\ref{tab:chordclass} show the frequencies of the chord 
types defined so far,  accounting for 77\% of the chords in the corpus. 
The no-chord type accounts for an additional 0.5\%, and in the following discussion, we detail how we map the remaining 22.5\% 
to the five types defined above.

\subsubsection{Triads}\label{subsec:triads}
Triads represent 16\% of the chords in the corpus, and as they do not contain a $7^{th}$ note, mapping them into the chord class types defined in the previous section can be indeterminate.  For example, a C major triad, consisting of the notes C, E, and G, shares all of its notes with both the C major7 and C dominant7 chords. 
We resolve triad ambiguities using principles from tonal harmony to identify whether they have a subdominant, dominant, or tonic function \cite{mulholland2013berklee}.

For a major triad, we look at the chord following it in the progression. If it has a root a fifth down and is a member of the major7 or minor7 classes we designate the triad as having a dominant function, and assign it to a dominant7 type with the same root.  Otherwise, it is assigned to a major7 class type.  We handle major triads with an added $9^{th}$ in the same way.

Augmented triads share their notes with dominant7$\sharp$5 chords, an alteration of the dominant.  We opted to map these to the dominant7 type class with the same root.  Finally, we map all the minor and diminished triads to their
corresponding minor7 and diminished7 types, respectively. 

\subsubsection{Sus Chords}\label{subsec:suschords}

Sus chords also have a harmonic function that depends on context \cite{levine2011jazz}. When followed by a dominant7 chord with the same root, they act like a subdominant and we opt to map them to a minor7 class with a root a fifth above.  For example, a G7sus4 would map to a Dm7.  Otherwise, they act like a dominant and we map it to a dominant7 class with the same root.

\subsubsection{Slash Chords}\label{subsec:slashchords}

Slash chords are chords played over a specific bass note, for example C/G or Dm7/G, where the symbol above (to the left of) the slash is a chord and below it a note.  If the bass note belongs to the chord above the slash (for example, C/G), it is an inversion.  For such case, we map it according to the chord above the slash.  

Slash chords are also commonly used to represent sus chords.  For example, Dm7/G is harmonically equivalent to G9sus4.  We map these according to the process for sus chords described in the previous section.  For all other slash chords, we map the chord as if the bass note were an extension or alteration of the chord above the slash. 

\subsubsection{Power Chords and Polychords}\label{subsec:powerchords}

Power chords consist of just two notes, a root and a fifth. As they have no 3$^{rd}$ or 7$^{th}$, they are harmonically ambiguous. With only 42 instances in our corpus, we have opted to map these chords to the \emph{no-chord} class.

With only 12 instances, polychords are rare.  These chords consist of either an upper and lower triad or triad and a $7^{th}$ chord.  We map polychords according to their lower structure, interpreting the upper structure as  extensions or alterations.  

\section{Key Signature Based Representation}\label{sec:keyest}

To make distributional semantics more effective, we are interested in a transposition invariant chord representation. 
Practices in music information retrieval have made use of both transposition to the same key (for example, \cite{tsushima2017function}) or to the same key signature (for example, \cite{mozer1994neural}). 
As knowing the key signature of a song is sufficient to transpose it, we opt for the latter, and transpose all songs to the key signature without sharps or flats (corresponding to C major/A minor). 

Transposing a song from one key signature to another presumes the former is known. The key signature listed in the database should be a credible source for this information. However, lead sheets and databases do not always accurately provide it.  Moreover, from extensive manual checking, we know that our own database contains several hundred songs for which the stated key signature is clearly in error (or for which we have doubts).
For this reason, we introduce a key signature estimation algorithm, which we describe in the following section.

\subsection{Key Signature Estimation Algorithm}\label{subsec:keyalg}

Several authors have proposed key estimation algorithms for various music information retrieval tasks \cite{mauch2009simultaneous,pauwels2014combining,rocher2010concurrent,noland2006key,benetos2014improving}. However, as discussed, our objective is not to estimate a song's key, but rather it's key signature. Some prior work also exists for key signature estimation \cite{foscarin2021pkspell}, however, it is based on machine learning models applied to MIDI data for classical music.  Here, we introduce a simple algorithm that requires very little computation, and estimates key signatures from symbolic jazz chord progressions.

Our algorithm estimates a song's key signature by selecting the one most consistent with its chords. Figure~\ref{fig:diatonic} is useful for illustrating our approach.  
The key signature in the figure has no sharps or flats, and so corresponds to the scale of C major. This scale generates the set of seven diatonic $7^{th}$ chords shown in the figure, and  results in major7, minor7, dominant7 and minor7$\flat$5 chord types.  These correspond to four of the five types discussed in Section~\ref{sec:modelorder}. The fifth type, diminished7 chords, is not associated to any major scale, and is not used in our estimation algorithm. 

For each chord in a progression, we map it to one of the 61 classes, as described in Section\ref{sec:modelorder}.  Except for chords from the diminished7 and no-chord classes, each chord is diatonically related to one or more key signatures and their related major scales.  The number of beats a chord is active is attributed to each one of the key-signatures to which it can belong.  After processing all the chords, the key signature accumulating the most beats is the resulting estimate for that song. 

\begin{figure}
 \centerline{\includegraphics[width=0.9\columnwidth]{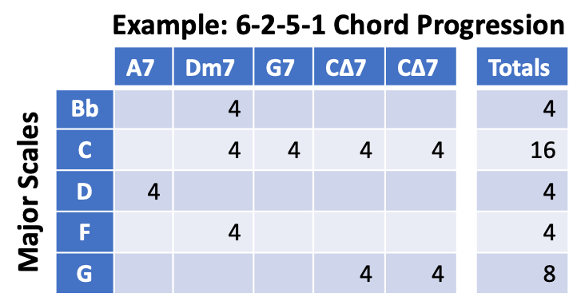}}
 \caption{Example application of the key signature estimation algorithm}
 \label{fig:keyestex}
\end{figure}

Figure~\ref{fig:keyestex} provides a concrete illustration of how the key estimation algorithm works for the case of a short chord progression: A7-Dm7-G7-CM7-CM7.  Each column of the table represents one measure, and in this example, there is one chord per measure.  The column labels correspond to the chords, and each row label is a key signature whose major scale diatonically contains one or more of the chords in the progression.  As shown, the A7 chord belongs to D major; the Dm7 chord belongs to Bb, C, and F major; G7 belongs to C major; and CM7 belongs to both C and G major.  Presuming four beats per measure, C  accumulates the most beats (16), and is the resulting key signature estimate for this short chord progression.

\subsection{Algorithm Evaluation}

As already mentioned, there are quite a few songs in our corpus where the key signature is incorrect or in doubt.  Nevertheless, it is worthwhile comparing the  results of our key estimation algorithm with the ones recorded in the corpus. Of the 2,612 songs, the algorithm concurs with the database for 1,763 (67.5\%) of them.  

\begin{figure}
 \centerline{\includegraphics[width=0.9\columnwidth]{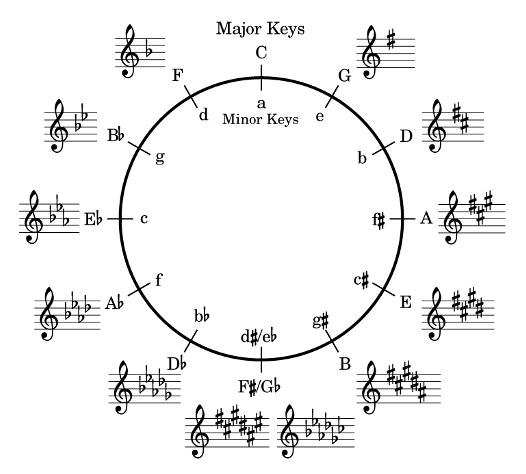}}
 \caption{Distances around the circle of fifths}
 \label{fig:circle_of_fifths}
\end{figure}

For the 849 songs with database keys signatures that don't agree with our estimates, we use the Circle of Fifths as a distance metric to evaluate the magnitude of differences between the two. Key signatures next to each other on the circle of fifths correspond to major scales that differ in a single note.  For example, as seen in Figure~\ref{fig:circle_of_fifths}, the key of F has the same notes as the adjacent key of C except for a B$\flat$, and G has the same notes as C except for an F$\sharp$.  

Table~\ref{tab:perfstats} recaps the statistics of 
applying the circle-of-fifths metric to estimated and database key signatures when evaluated for all of the songs in the corpus. The left-most column in the figure indicates the distance in number of sharps or flats between the estimated and database key.  The row with the distance of $0$ corresponds to agreement.  As already observed, a total of 1,763 cases fall into that category. There are 304 cases of songs where the estimate is 1$\flat$ away from the database key signature, and 183 that are 1$\sharp$ away.  The number of remaining cases for larger circle-of-fifths distances tail off rapidly.

\begin{table}
\centering
\begin{tabular}{c|r|r}
\textbf{Distance} & \textbf{Number} & \textbf{Percent} \\
\hline \hline
6$\flat$       & 10                         & 0.4\%                       \\
5$\flat$       & 22                         & 0.8\%                       \\
4$\flat$       & 33                         & 1.3\%                       \\
3$\flat$       & 55                         & 2.1\%                       \\
2$\flat$       & 99                         & 3.8\%                       \\
1$\flat$       & 304                        & 11.6\%                      \\
0        & 1763                       & 67.5\%                      \\
1$\sharp$       & 183                        & 7.0\%                       \\
2$\sharp$       & 22                         & 0.8\%                       \\
3$\sharp$       & 25                         & 1.0\%                       \\
4$\sharp$       & 12                         & 0.5\%                       \\
5$\sharp$       & 1                          & 0.0\%                       \\
Ambig.   & 123                        & 4.7\%                      
\end{tabular}
\caption{Key signature estimation statistics}
\label{tab:perfstats}
\end{table}

The last row of Table~\ref{tab:perfstats} is labelled "Ambig.".   There are 123 songs in the database for which the key estimation algorithm returns a non-unique result, and finds two or more major scales that accumulate the most beats. This represents 4.7\% of the songs in the corpus, and when this occurs our estimation algorithm defaults to the database key.  

\subsection{Mapping to Scale Position and Chord Classes}

Once all the chords in the corpus have been attributed to their respective classes, and the key estimation algorithm applied to each song, all songs can be transposed to a common key signature. 
We selected the key signature of C major / A minor, but that choice has no influence on the calculation of the co-occurrence matrix or the similarity metric. 
The combination of a scale position and chord type defines a chord class. 
The combination of our 5 chord types with 12 scale positions plus the \emph{no-chord} symbol leads to 61 chord classes.   

The chord classes can be mapped to Roman numeral notation, which is commonly used in harmonic theory. 
The choice of numerals depends on the assumed root note (which determines the mode). 
Table~\ref{tab:map2roman} shows the mapping for C major. 
As an example for this key signature, a sequence of chords such as A7-Dm-G7-CM maps to vi7-iim-v7-iM. 
Under our schema, the relative minor cadence  Bh-E7-Am maps to viih-iii7-vim. 

\begin{table}
\centering
\begin{tabular}[t]{c|c}
\textbf{Root} & \textbf{Roman Numerals (C major)} \\
\hline \hline
C    & i     \\
D$\flat$   & $\flat$ii   \\
D    & ii    \\
E$\flat$   & $\flat$iii  \\
E    & iii   \\
F    & iv    \\
G$\flat$   & $\flat$v    \\
G    & v     \\
A$\flat$   & $\flat$vi   \\
A    & vi    \\
B$\flat$   & $\flat$vii  \\
B    & vii  
\end{tabular}
\caption{Roman numeral notation for C major}
\label{tab:map2roman}
\end{table}

\section{Co-Occurrence Matrix}

The model-reduced, transposed versions of songs in our corpus can be used to compute a co-occurrence matrix \cite{bordag2008comparison,globerson2004euclidean,leydesdorff2006co,lund1996producing}. 
Prior to the computation, we introduce a \emph{<START>} and \emph{<END>} symbol at the beginning and end of each song's chord progression.  
This increases the total number of classes to 63. 
Using a symmetric context window of size 1, the $ij^{th}$ element of the co-occurrence matrix represents the number of times the $j^{th}$ chord class occurs in the context of the $i^{th}$ chord class.
The elements of the co-occurrence matrix, $\mathbf{C}_{ij}$, are obtained by tabulating the number of times the $j^{th}$ chord class occurs next to the $i^{th}$ class when scanning through all the chord progressions in the corpus.  Given a corpus of size $D$, this can be expressed as 
\begin{equation}
\mathbf{C}_{ij} = 
\sum_{k=1}^{D} \sum_{i=1}^{N_k} \begin{cases} 1, & \text{if $s_{k_{(i-1)}} = j$ and $(i-1) > 0$} \\ 
                                            1, & \text{if $s_{k_{(i+1)}} = j$ and $(i+1) \leq N_k$} \\
                                            0, & \text{otherwise} \end{cases}
\label{equ:cooccur}
\end{equation}
where 
$s_{k_1}, \ldots, s_{k_{N_k}}$
is the chord class sequence for song $k$, and $i,j \in \{1, \ldots ,63\}$ are the corresponding chord class indices. 
The rows $\mathbf{C}_i $ 
of the matrix can be 
used as a vector embedding for the chord classes.  
Because co-occurrence matrices capture contextual information, the vectors of chord classes that have similar harmonic function are expected to be close to each with respect to the cosine similarity measure.

Table~\ref{tab:cosine_distance} illustrates two examples of how the co-occurrence matrix in Equation ~\ref{equ:cooccur} captures semantic structure in the chord progressions from our corpus.  
The left-hand table shows the five closest chord classes to $\flat$ii7.  The lowest row of the table is the cosine similarity of this chord with itself and, as expected, has a value of 1. Of the 62 remaining chord classes, the next closest in cosine similarity is the v7 chord class with a value of 0.821. 
From a music theory perspective, this result seems reasonable because the $\flat$ii7 and v7 are tritone substitutes that are often used as reharmonizations for each other.  

\begin{table}
\centering
\begin{tabular}{l|r}
\textbf{Class}                 & \textbf{bii7} \\
\hline \hline
iio                            & 0.751         \\
\textless{}START\textgreater{} & 0.781         \\
io                             & 0.811         \\
$\flat$iio                           & 0.813         \\
v7                             & 0.821         \\
$\flat$ii7                           & 1.000        
\end{tabular}
\quad \quad
\begin{tabular}{l|l}
\textbf{Class}               & \textbf{iim} \\
\hline \hline
v7                           & 0.786        \\
\textless{}END\textgreater{} & 0.829        \\
vo                           & 0.850        \\
$\flat$vi7                         & 0.914        \\
iih                          & 0.924        \\
iim                          & 1.000       
\end{tabular}
\caption{Cosine distance and harmonic similarity for the
$\flat$ii7 and the iim chord classes}
\label{tab:cosine_distance}
\end{table}

The table on the right-hand side of Table~\ref{tab:cosine_distance} shows the top five chord classes nearest to iim.  The closest, iih,  seems satisfactory from a music theory perspective. 
The iim is the most common subdominant chord, and is the first chord in the major cadence, iim-v7-iM.  
The iih is a common reharmonization for the iim, and is also the first chord in the minor cadence for the parallel minor.

\section{Membrane-Area Distance Metric}

The normalized chord vectors derived from the co-occurrence matrix can be used to plot the path of a song's progression through 63-dimensional space. Starting from the origin, the sequence of chord vectors can be concatenated from head to tail, beginning with the <START>, and terminating with the <END> vector. Each unit vector is scaled by the number of beats of the chord it represents, and the result is a piecewise linear function through $\mathbf{R}^{63}$. 

The piecewise linear functions for two identical chord progressions would, naturally, overlay each other, and two harmonically similar songs should trace similar paths through $\mathbf{R}^{63}$.  

Expressed formally, we represent song vector paths by piecewise linear functions of the form $\bm{f}(t) \in \mathbf{R}^{63}$, where $t \in [0,1]$ is a parametric variable representing the number of normalized beats traversed in the song. We can move along the entire length of $\bm{f}$ in discrete, equal increments, $dt$, where the starting point of the function, $\bm{f}(0)$ at $t = 0$ is the origin, and the end point of the function is at $t=1$. 

Given two songs and their corresponding piecewise linear functions,  $\bm{f}(t)$ and $\bm{g}(t)$, and letting $N = 1/dt$, we can define a distance metric between them as the area of a 2D membrane, $M$, stretched between the two paths.  $M$ is calculated as the integral obtained in the limit of
\begin{equation}\label{eq:distance}
M(\bm{f},\bm{g}) = \lim_{dt \rightarrow 0}\sum_{n = 0}^{N} \|\bm{f}(n dt) - \bm{g}(n dt)\|dt
\end{equation}
where $\|\cdot\|$ is the Euclidean norm. 

\begin{figure}
 \centerline{\framebox{\includegraphics[width=0.9\columnwidth]{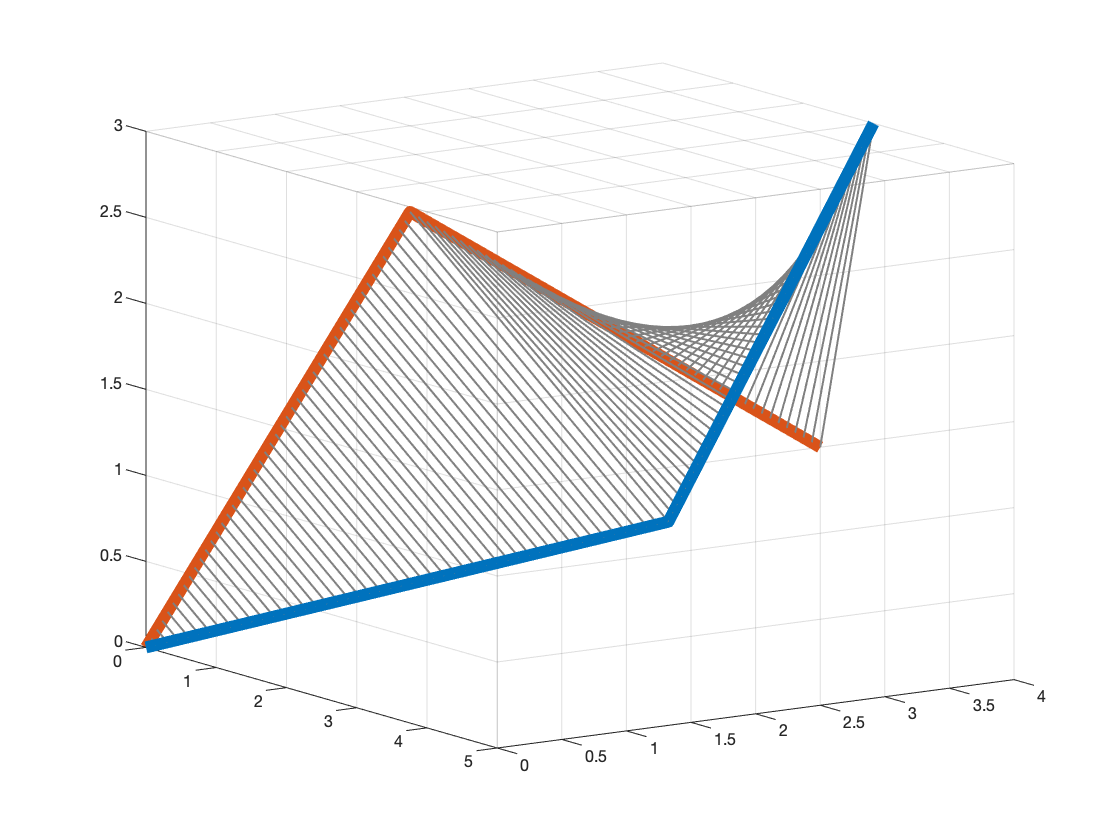}}}
 \caption{Conceptual illustration of the membrane-area distance metric}
 \label{fig:membrane}
\end{figure}

Figure~\ref{fig:membrane} is a graphical illustration of how the measure in Equation\ref{eq:distance} is evaluated.  The red and blue lines represent two different songs, each having two chords (excluding the terminal symbols).  The songs begin at the origin, and the chord vectors are added head-to-tail. The metric is approximated by summing the lengths of the $N$ equally spaced black line segments drawn between the two songs.

\section{Experiment}

We have found no scholarly reference data sets for contrafacts that can be used as ground truth for a comprehensive evaluation of our method.
Although there is a list of 252 jazz contrafacts assembled in a Wikipedia article \cite{wiki2018list}, only 91 of those songs are found in our corpus.  Moreover, quite a few of the examples cited in the article are not full contrafacts, meaning that the song only borrows a portion of the harmony from another song. The eight bars of the \emph{Rhythm Changes} bridge is an example of this.  

Despite the lack of comprehensive ground truth,
we can illustrate the performance and characteristics of our method with an example.  
As the membrane-area metric can be computed for any pair of songs, it can be used to search for contrafacts. If $f^*$ is the vector path of a reference song and $S$ is the set of all songs in the corpus excluding $f^*$ the song, $\hat{f}$, harmonically closest to $f^*$ is

\begin{equation}\label{equ:optimize}
 \hat{f} = \arg \min_{f \in S} M(f^*, f)
\end{equation}

\begin{figure}
 \centerline{\includegraphics[width=0.95\columnwidth]{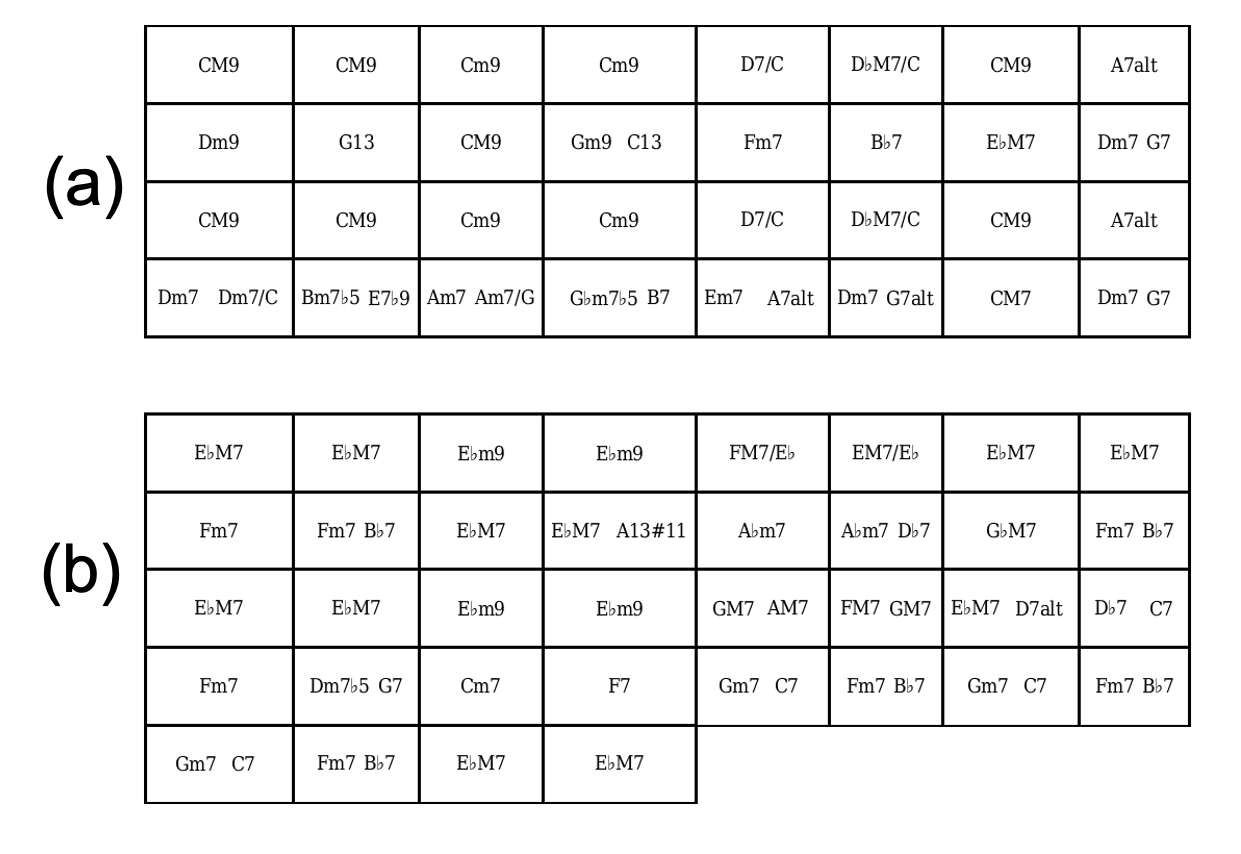}}
 \caption{Chord progressions for (a) \emph{On Green Dolphin Street}, and (b) \emph{Green St. Caper}}
 \label{fig:GreenDolphinSt}
\end{figure}

As an example, Figure~\ref{fig:GreenDolphinSt}(a) illustrates the chord progression for the jazz standard \emph{On Green Dolphin Street}. Evaluating the distance between this song and every other one in the corpus, the closest is the well-known contrafact, \emph{Green St. Caper} \cite{wiki2018list}, shown in Figure~\ref{fig:GreenDolphinSt}(b).

\begin{figure}
 \centerline{\includegraphics[width=0.95\columnwidth]{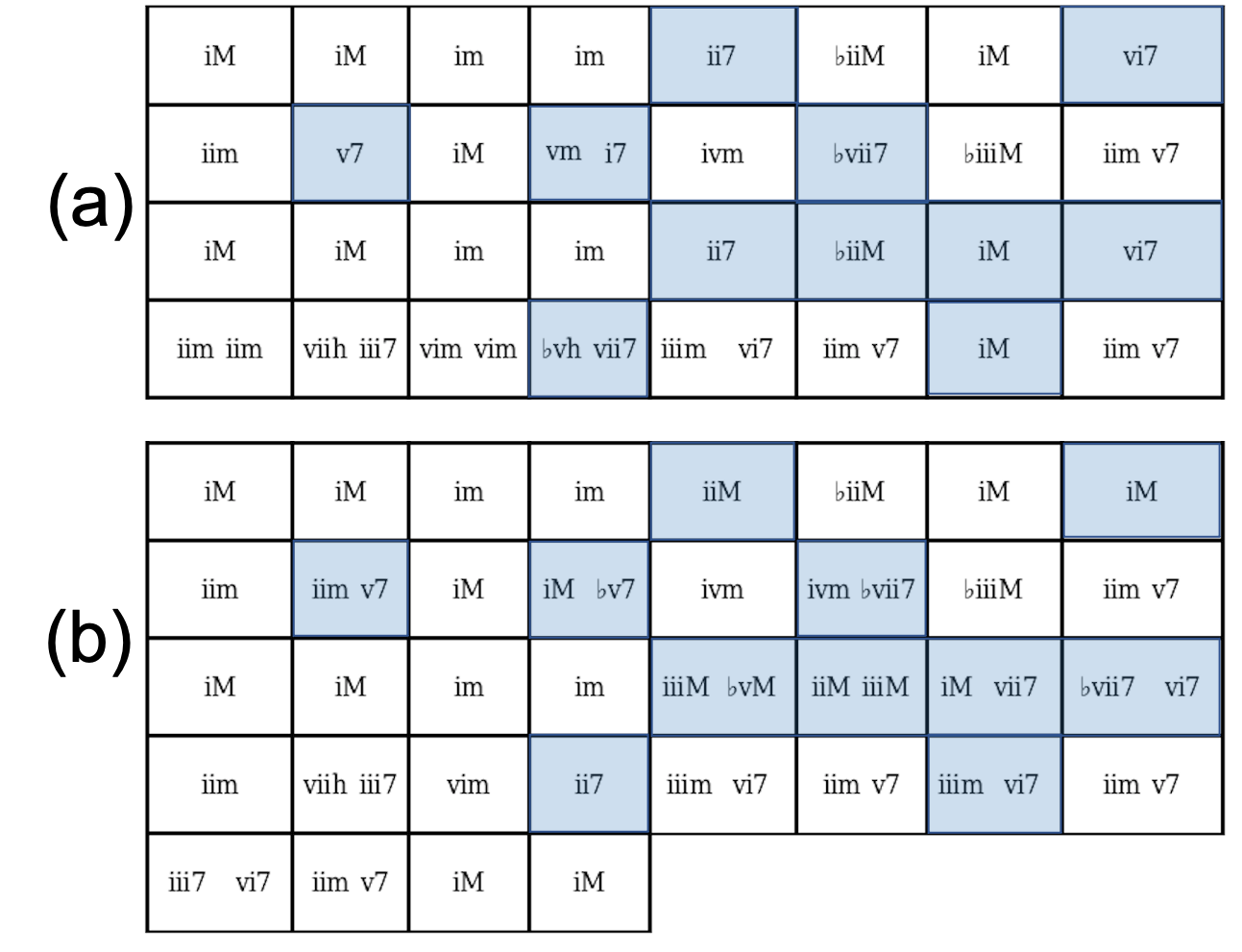}}
 \caption{Roman numeral notation for (a) \emph{On Green Dolphin Street}, and (b) \emph{Green St. Caper}.  Measures with blue backgrounds are variations in harmonization.}
 \label{fig:Compare}
\end{figure}

Although contrafacts, these two songs have numerous differences. \emph{On Green Dolphin Street} is in the key of C, and \emph{Green St. Caper} is in E$\flat$.  The former has 32 bars, and the latter has 36.  There are also numerous differences in harmonization.  To more clearly see this last point, it is easier to examine the two songs in Roman numeral notation, as shown in Figure~\ref{fig:Compare}. 

There are 10 measures in the two songs that have different chords (highlighted with pale blue backgrounds). For example, measure 12 contains a vm-i7 in the reference song and iM-$\flat$v7 in the contrafact. Here the contrafact has replaced the vm by extending the duration of the iM from the previous bar, and has replaced the i7 with its tritone substitute, the $\flat$v7. This is a concrete instance of the example discussed in Table~\ref{tab:cosine_distance}. The differences seen in each of the other highlighted measures correspond to well-known  reharmonizations as discussed in \cite{berkman2013jazz}.

\section{Conclusions and Future Work}

This paper makes several contributions. The first is the use of music theory to reduce the dimensionality of chord space.
Our method is comprehensive, detailing how to map every one of the 1,542 chord types found in our corpus to 61 classes.

The resulting chord class progressions are used to compute a dense co-occurrence matrix without needing to resort to non-parametric approximations such as truncated SVD or gradient descent.  Furthermore, we show that our corpus generates a  co-occurrence matrix that is able to capture semantic information about harmony.  Using the cosine similarity measure, we present examples illustrating that rows of the co-occurrence matrix embody characteristics of common reharmonizations.  

Using the normalized rows of the matrix as vector embeddings of chord classes, we modeled songs as piecewise linear paths in $\mathbf{R}^{63}$.  A novel distance metric, the membrane-area, was introduced, and used as a measure of harmonic similarity between songs.  An example of its application was presented, successfully finding a song's known contrafact even though the two songs have numerous differences in harmonization. 

This paper examines one type of vector embedding for chords, but other approaches such as TF-IDF \cite{ramos2003using} or Word2Vec \cite{mikolov2013efficient} are worthwhile exploring.  The latter is particularly interesting because it has been shown to capture additional semantic characteristics (for example, analogy) when used in natural language processing. 

The vector embeddings discussed in this paper could be used as inputs to machine learning architectures such as RNNs \cite{chung2015recurrent} and Transformers \cite{vaswani2017attention}.  This could be a promising avenue of application for MIR tasks such as automatic chord recognition \cite{pauwels201920}, automatic music transcription \cite{benetos2018automatic}, genre detection \cite{scaringella2006automatic}, and cover song identification \cite{yesiler2020accurate,xu2018key,khadkevich2013large,bertin2011large}.

\bibliography{MyRefs}

\end{document}